\documentclass[a4paper, 12pt]{article}
\author{C. J. de Matos\footnote{ESA-HQ, European Space Agency, 8-10 rue Mario Nikis, 75015 Paris, France, e-mail: Clovis.de.Matos@esa.int}
}
\title{Approximation to the Second Order Approximation of Einstein Field Equations with a Cosmological Constant in a Flat Background}

\begin{document}

\maketitle \begin{abstract} Einstein field equations with a
cosmological constant are approximated to the second order in the
perturbation to a flat background metric. The final result is a
set of Einstein-Maxwell-Proca equations for gravity in the weak
field regime. This approximation procedure implements the breaking
of gauge symmetry in general relativity. A brief discussion of the
physical consequences (Pioneer anomalous deceleration) is proposed
in the framework of the gauge theory of gravity.
\end{abstract}

\section{Introduction}
Einstein introduced by hand a Cosmological Constant (CC) $\Lambda$
in his field equations to compensate for the universe expansion,
and predict a stationary universe. He made reference to the CC as
being his biggest mistake, when Hubble discovered that our
universe was in expansion. Modern observational cosmology reveals
a flat universe in accelerated expansion, which might be explained
through a positive CC different from zero. What Einstein
considered to be his worst idea, might well become its smartest
thought!

The only homogeneous and isotropic vacuum solution of the Einstein
field equations (EFE) with CC is (anti) de Sitter space, not
Minkowski space \cite{Higuchi}. However, since contemporary
observational cosmology tend to measure a flat universe in
accelerated expansion \cite{Spergel}, although Minkowski metric is
not a solution of the vacuum EFE with CC, in the present work we
impose the \emph{ansatz} of a flat metric for the background
instead of the (anti) de Sitter background metric, and we apply an
approximation procedure, closely related to the standard
linearization method \cite{de Matos1} \cite{Peng} \cite{Campbell},
leading to "massive" GravitoElectroMagnetism (GEM). The result is
a set of Einstein-Maxwell-Proca type equations for GEM. The
\emph{ansatz} of a flat background metric to approximate EFE with
a CC instead of the (anti) de Sitter background corresponds to the
implementation of a spontaneous breaking of gauge invariance in
general relativity \cite{de Matos2}. As we will see in the
conclusion, the Gauge Theory of Gravity (GTG) is particularly well
suited to discuss the physical consequences of this symmetry
breaking \cite{Hestenes}.
\section{Second Order Approximation of Einstein Field Equations with
a Cosmological Constant} In the following we propose an
approximation of the second order approximation of Einstein Field
Equations with cosmological constant. The following assumptions
underly the approximation procedure:

\begin{enumerate}
\item  \label{approx_1}the mass densities are normal (no dwarf
stars), and correspond to \emph{local} physical systems located in
the Earth laboratory or in the solar system.

\item  \label{approx_2}All motions are much slower than the speed of light,
so that special relativity can be neglected. (Often special
relativistic effects will hide general relativistic effects),
$v<<c$.

\item  \label{approx_3}The kinetic or potential energy of all the bodies
being considered is much smaller than their mass energy,
$T_{\mu\nu}<<\rho c^2$.

\item  \label{approx_4}The gravitational fields are always weak enough so
that superposition is valid, $\phi<<c^2$.

\item  \label{approx_5}The distances between objects is not so large that we
have to take retardation into account. (This can be ignored when
we have a stationary problem where the fields have already been
prescribed and are not changing with time.)

\item  \label{approx_55}We consider a running cosmological
constant which depends on the local density of mass of the
physical system being considered
\[
\Lambda=\frac{4\pi G}{c^2}\rho
\]

\item \label{approx 6}We consider that the proposed approximation is only valid in the following
range of distances:
\[
\sqrt{|\frac{h_{\alpha\beta}}{\Lambda}|}<<r<<\sqrt{|\frac{1}{\Lambda}|}
\]
defining the characteristic length scale for the physical system
being considered. This restriction allows to neglect second order
terms in the perturbation to Minkowsky's metric
$|h_{\alpha\beta}|^2$, but we do not neglect terms involving
simultaneously the perturbation to the metric and the cosmological
constant, $|\Lambda h_{\alpha\beta}|$.

\item \label{approx 7} The approximated second order EFE are solved, by approximation, using
the solutions for the perturbation, $|h_{\alpha\beta}|$, to
Minkowsky's metric obtained in the linear approximation, in which
only first order terms in $|h_{\alpha\beta}|$ are considered.
\end{enumerate}

We start with Einstein Field Equations (EFE)with a cosmological
constant.
\begin{equation}
R_{\alpha\beta}-\frac{1}{2}g_{\alpha\beta}R-\Lambda
g_{\alpha\beta}=\frac{8\pi G}{c^4} T_{\alpha\beta}\label{equ1}
\end{equation}
The weak field approximation assumes small perturbations,
$|h_{\alpha\beta}|<<1$, of Minkowsky's metric
$\eta_{\alpha\beta}(+---)$ (Landau-Lifschitz "timelike
convention"). This approximation is deliberately kept also for the
case of having a cosmological constant different from zero.
\begin{equation}
g_{\alpha\beta}\approx\eta_{\alpha\beta}+h_{\alpha\beta}\label{equ2}
\end{equation}
Doing Equ.(\ref{equ2}) into Equ.(\ref{equ1}) with the derivation
indices obeying the same rule as the covariant indices,
$f^{,\mu}=\eta^{\mu\nu}f_{,\nu}$, we obtain:
\begin{equation}
-\frac{1}{2}\Big(\bar{h}^{,\mu}_{\alpha\beta,\mu}+\eta_{\alpha\beta}\bar{h}^{,\mu\nu}_{\mu\nu}-\bar{h}^{,\mu}_{\alpha\mu,\beta}-\bar{h}^{,\mu}_{\beta\mu,\alpha}\Big)
-\Lambda\Big(\eta_{\alpha\beta}+h_{\alpha\beta}\Big)=\frac{8\pi
G}{c^4} T_{\alpha\beta}\label{equ3}
\end{equation}
As usual in order to simplify the linearization procedure we have
introduced the intermediate tensor:
\begin{equation}
\bar{h}_{\alpha\beta}=h_{\alpha\beta}-\frac{1}{2}\eta_{\alpha\beta}h\label{equ4}
\end{equation}
where
$h=h^{\mu}_{\mu}=\eta^{\mu\nu}h_{\mu\nu}=h_{00}-h_{11}-h_{22}-h_{33}$
is the trace of the perturbation tensor. Imposing the harmonic
gauge condition
\begin{equation}
\bar{h}^{\mu\nu}_{,\nu}=0\label{equ5}
\end{equation}
Equ.(\ref{equ3}) reduces to
\begin{equation}
\bar{h}^{,\mu}_{\alpha\beta,\mu}+2\Lambda\Big(\eta_{\alpha\beta}+h_{\alpha\beta}\Big)=-\frac{16\pi
G}{c^4} T_{\alpha\beta}\label{equ6}
\end{equation}
Equ.(\ref{equ6}) can be written in function of the Dalembertian
operator, $\triangle$. If $f$ is a given function, then
\begin{equation}
\triangle{f}=f^{,\mu}_{,\mu}=\eta^{\mu\nu}f_{,\mu\nu}=\Bigg(\frac{\partial^2}{(\partial
x^0)^2}-\frac{\partial^2}{(\partial x^i)^2}\Bigg)f\label{equ7}
\end{equation}
Where $x_0=ct$. Therefore Equ.(\ref{equ6}) becomes
\begin{equation}
\triangle\bar{h}_{\alpha\beta}+2\Lambda\Big(\eta_{\alpha\beta}+h_{\alpha\beta}\Big)=-\frac{16\pi
G}{c^4} T_{\alpha\beta}\label{equ8}
\end{equation}
This is the approximated second order form of EFE with a
cosmological constant, assuming a flat background for the metric.
We will now solve these equations, by approximation, using the
solutions of the perturbation to Minkowsky's metric we obtain in
the case of linear EFE without CC.

To split spacetime into gravitoelectric and gravitomagnetic parts
we consider respectively the energy-momentum tensor components:
\begin{equation}
T_{00}=\rho c^2,\label{equ9}
\end{equation}
and
\begin{equation}
T_{0i}=-\rho c v_i.\label{equ10}
\end{equation}
The solution of EFE without cosmological constant (with a flat
background), $\triangle\bar{h}_{00}=-\frac{16\pi G}{c^4} T_{00}$,
for the energy momentum tensor component given by
Equ.(\ref{equ9})is:
\begin{equation}
h_{00}=\frac{2\phi}{c^2}\label{equ11}
\end{equation}
Where $\phi$ is the gravitational scalar potential. The solution
of EFE without cosmological
constant,$\triangle\bar{h}_{0i}=-\frac{16\pi G}{c^4} T_{0i}$, for
the energy momentum tensor component of Equ.(\ref{equ10}) is:
\begin{equation}
h_{0i}=-\frac{4A_{gi}}{c}.\label{equ12}
\end{equation}
Where $A_{gi}$ are the three components of the gravitomagnetic
vector potential.

Writing the Einstein tensor in function of the intermediate tensor
$\bar{h}_{\alpha\beta}$, and using the gauge condition of
Equ.(\ref{equ5}), Einstein tensor reduces to the tensor
$G_{\alpha\beta\mu}$.
\begin{equation}
G_{\alpha\beta\mu}=\frac{1}{4}\Bigg(\bar{h}_{\alpha\beta,\mu}-\bar{h}_{\alpha\mu,\beta}\Bigg)\label{equ13}
\end{equation}
Using Equ.(\ref{equ13}) one can re-write Equ(\ref{equ8}) under the
following form:
\begin{equation}
\frac{\partial G_{\alpha\beta\mu}}{\partial
x^{\mu}}+\frac{1}{2}\Lambda\Big(\eta_{\alpha\beta}+h_{\alpha\beta}\Big)=-\frac{4\pi
G}{c^4} T_{\alpha\beta}\label{equ14}
\end{equation}
We can also use the tensor Equ.(\ref{equ13}) to express the
gravitational field:
\begin{equation}
g_i=-c^2G_{00i}.
\end{equation}
Which can also be written in terms of the gravitational scalar
potential $\phi$ and of the gravitomagnetic vector potential
$\vec{A_g}$.
\begin{equation}
\vec{g}=-\bigtriangledown \phi - \frac {\partial
\vec{A_g}}{\partial t}
\end{equation}

Similarly we formulate the gravitomagnetic field as follows:
\begin{equation}
cG_{0ij}=-(A_{gi,j}-A_{gj,i})
\end{equation}
which obviously shows that the gravitomagnetic field $\vec{B_g}$
is generated by a vectorial potential vector $\vec{A_g}$.
\begin{equation}
\vec{B_g}=\nabla\times \vec{A_g}\label{Bg}
\end{equation}
We have now everything we need to derive Proca-type equations for
gravity. For the energy momentum tensor component of $T_{00}$ of
Equ.(\ref{equ9}), Equ.(\ref{equ14}) reduces to:
\begin{equation}
\frac{\partial G_{00\mu}}{\partial
x^{\mu}}+\frac{1}{2}\Lambda\Big(\eta_{00}+h_{00}\Big)=-\frac{4\pi
G \rho}{c^2}\label{equ15}
\end{equation}
Substituting the solution Equ(\ref{equ11}) of EFE, into
Equ.(\ref{equ15}) we can approximate the divergent part of the
gravitational field:
\begin{equation}
\nabla\cdot\vec{g}=-4\pi G
\rho-\Lambda\phi-\frac{1}{2}c^2\Lambda\label{equ16}
\end{equation}
For the energy momentum tensor component $T_{0i}$ of
Equ.(\ref{equ10}), Equ.(\ref{equ14}) reduces to:
\begin{equation}
\frac{\partial G_{0i\mu}}{\partial
x^{\mu}}+\frac{1}{2}\Lambda\Big(\eta_{0i}+h_{0i}\Big)= \frac{4\pi
G}{c^3} \rho v_i \label{equ17}
\end{equation}
Substituting the solution Equ(\ref{equ12}) of EFE, into
Equ.(\ref{equ17}) we can approximate the rotational part of the
gravitomagnetic field:
\begin{equation}
\nabla\times\vec{B_g}=-\frac{4\pi G}{c^2}
\vec{j_m}+\frac{1}{c^2}\frac{\partial\vec g}{\partial t}-2 \Lambda
\vec{A_g}\label{equ18}
\end{equation}
Where $\vec{j_m}=\rho\vec{v}$ is the mass current. The tensor
$G_{\alpha\beta\mu}$, Equ.(\ref{equ13}), has the following
property:
\begin{equation}
G^{\alpha\beta\mu,\lambda}+G^{\alpha\lambda\beta,\mu}+G^{\alpha\mu\lambda,\beta}=0.\label{19}
\end{equation}
which are equivalent to the two other set of Maxwell like
equations for gravity,
\begin{equation}
\nabla\cdot\vec{B_g}=0\label{equ20}
\end{equation}
and
\begin{equation}
\nabla\times\vec g=-\frac{\partial \vec{B_g}}{\partial
t}\label{equ21}
\end{equation}
Note also that Equ.(\ref{equ20}) is a direct and trivial corollary
of the definition of the gravitomagnetic field Equ.(\ref{Bg}). As
we see, Equs. (\ref{equ20}) and (\ref{equ21}) are not affected by
the cosmological constant.

In summary Equs (\ref{equ16}) (\ref{equ18}) (\ref{equ20}) and
(\ref{equ21}) form a set of Einstein-Maxwell-Proca equations for
gravity in the weak field regime:
\begin{eqnarray}
\nabla\cdot\vec{g}&=& -4\pi G \rho-\Lambda\phi-\frac{1}{2}c^2\Lambda \label{equ22}\\
\nabla\cdot\vec{B_g}&=& 0 \label{equ23}\\
\nabla\times\vec g &=& -\frac{\partial \vec{B_g}}{\partial t} \label{equ24}\\
\nabla\times\vec{B_g} &=& -\frac{4\pi G}{c^2}
\vec{j_m}+\frac{1}{c^2}\frac{\partial\vec g}{\partial t}-2\Lambda
\vec{A_g}\label{equ25}
\end{eqnarray}
These equations are closely analogous to the ones derived by
Argyris to investigate the consequences of massive gravitons in
general relativity \cite{Argyris}.

Considering the case of an universe empty of material
gravitational sources, $\rho=0$ and $\phi=0$, Equ.(\ref{equ22})
reduces to:
\begin{equation}
\nabla\vec g=-\frac{1}{2}c^2\Lambda \label{equ27}
\end{equation}
Integrating this equation over a volume bounded by a sphere of
radius $R$ we obtain a fundamental "Machian-type" accelerated
contraction of that volume, which only depends on its radius and
on the value of the CC, $\Lambda$.
\begin{equation}
g=\frac{1}{6} c^2 \Lambda R \label{equ28}
\end{equation}
This acceleration is directed inwards on the boundary of the
sphere. Doing the radius of the observable universe expressed in
function of the CC

$R=R_U=\sqrt{3/\Lambda}$ and using $\Lambda=1.29\times10^{-52}
[m^{-2}]$ derived from the value of $71 [Km.S^{-1}.Mpc^{-1}]$ for
the Hubble constant $H=c\sqrt{\Lambda/3}$ assumed by Nottale
\cite{Nottale}, into Equ.(\ref{equ28}) we obtain a fundamental
cosmic deceleration.
\begin{equation}
g=\frac{1}{2\sqrt 3} c^2 \Lambda^{1/2}=2.9\times10^{-10} m.s^{-2}
\label{equ29}
\end{equation}
So the linearized acceleration is a deceleration,it is interesting
to note that this value is in good agreement with the Pioneer
anomalous deceleration, whose current measured value is
$a_{Pio}=(8.5\pm1.3)\times10^{-10} m.s^{-2}$\cite{Pioneer}. If
this is a correct contribution to the Pioneer anomaly, this would
imply that this deceleration would be independent of the origin at
which the physical observer, measuring the Pioneer deceleration,
is located. The deceleration should always be directed towards the
observer, because the universe has no preferred center, and the
deceleration only depends on the universe radius. Similarly the
acceleration of expansion of the universe is independant of the
origin at which the astronomer measuring it is located. The
Pioneer anomalous deceleration would be a kind of local
"Machian-type back reaction"to the accelerated cosmological
expansion.

It might seem odd that the linearized analysis predicts an inward
acceleration $\vec{g}$ when $\Lambda >0$, as a positive value of
$\Lambda $ is known to cause the universe to accelerate outwards.
The acceleration in the second case is not $\vec{g}$ as expressed
in Equ.(\ref{equ28}), but the acceleration $\ddot{R}/R$ of the
scale factor in the Friedmann Robertson Walker (FRW) metric:
\[
ds^{2}=c^{2}d\tau ^{2}-R^{2}(\tau )(dx^{2}+dy^{2}+dz^{2})
\]
According to the Friedmann equation (the Einstein equation for
$R_{00}$, with non-negligible isotropic pressure $p$),
\[
\ddot{R}/R=\frac{1}{3}c^{2}\Lambda -\frac{4}{3}\pi G(\rho
+3pc^{-2})
\]
and so a positive $\Lambda $ increases the value of $\ddot{R}/R$.
The two accelerations $g$ and $\ddot{R}/R$ are not directly
comparable, as the FRW\ metric is not written in harmonic
coordinates.
\section{Spontaneous Breaking of Gauge Invariance in General Relativity}
General Relativity is founded on the \emph{principle of
equivalence}, which rests on the equality between the inertial and
the gravitational mass of any physical system, and formulates that
\emph{at every space-time point in an arbitrary gravitational
field it is possible to choose a "locally inertial coordinate
system" such that, within a sufficiently small region of the point
in question, the laws of nature take the same form as in
unaccelerated Cartesian coordinate systems in the absence of
gravity}. In other words, The inertial frames, that is, the
"freely falling coordinate systems", are indeed determined by the
local gravitational field, which arises from all the matter in the
universe, far and near. However, once in an inertial frame, the
laws of motion are completely unaffected by the presence of nearby
masses, either gravitationally or in any other way.

Following Steven Weinberg, the \emph{Principle of General
Covariance} (PGC) is an alternative version of the principle of
equivalence\cite{Weinberg}, which is very appropriate to
investigate the field equations for electromagnetism and
gravitation. It states that \emph{a physical equation holds in a
general gravitational field, if two conditions are met}:
\begin{enumerate}
\item The equation holds in the absence of gravitation; that is, it
agrees with the laws of special relativity when the metric tensor
$g_{\alpha\beta}$ equals the Minkowsky tensor $\eta_{\alpha\beta}$
and when the affine connection $\Gamma_{\beta\gamma}^{\alpha}$
vanishes.
\item The equation is generally covariant; that is, it preserves
its form under a general coordinate transformation $x \rightarrow
x'$.
\end{enumerate}

It should be stressed that general covariance by itself is empty
of physical content. The significance of the principle of general
covariance lies in its statement about the effects of gravitation,
that a physical equation by virtue of its general covariance will
be true in a gravitational field if it is true in the absence of
gravitation. The PGC is not an invariance principle, like the
principle of Galilean or special relativity, but is instead a
statement about the effects of gravitation, and about nothing
else. In particular general covariance does not imply Lorentz
invariance. Any physical principle such as the PGC, which takes
the form of an invariance principle but whose content is actually
limited to a restriction on the interaction of one particular
field, is called a dynamic symmetry. Local gauge invariance, which
governs the electromagnetic interaction is another important
dynamical symmetry. We can actually say that the Principle of
General Covariance in general relativity is the analogous of the
Principle of Gauge Invariance in electrodynamics. Spontaneous
breaking of gauge invariance in general relativity would therefore
correspond to a breaking of the PGC.

In contrast to the Einstein-Maxwell type theory of linear
gravitation \cite{Peng}, in the Einstein-Maxwell-Proca type
theory, Equs.(\ref{equ22})-(\ref{equ25}), the potentials $\phi$
and $\vec A_g$ are directly measurable quantities so that gauge
invariance is not possible, and the Lorentz gauge condition
\begin{equation}
\nabla.\vec  A_g + \frac {1}{c} \frac{\partial \phi}{\partial
t}=0\label{equ26}
\end{equation}
is required in order to conserve energy \cite{Argyris}. Since
$\Lambda\neq0$ is not consistent with gauge invariance, Proca
generalisation of gravitoelectromagnetism could be aesthetically
defective in the eyes of many theoretical physicists. However, the
only certain statements about the value of $\Lambda$ that can be
made must be based on experiment, and cosmological observations.

It should be noted that Einstein-Maxwell-Proca equations,
Equs.(\ref{equ22})-(\ref{equ25}), form a good phenomenological
base to investigate the GEM properties of superconductors \cite{de
Matos2}. They offer also an interesting perspective on Mach's
principle as formulated in the framework of relational
mechanics\cite{Assis}.
\section{Conclusion}
The set of Einstein-Maxwell-Proca equations derived from EFE with
CC assuming a flat background, can be understood as the result of
spontaneous breaking of gauge invariance in general relativity,
which is physically revealed through the violation of the
principle of general covariance. This is elegantly expressed in
the framework of gauge theory of gravity. The Gauge Theory of
Gravity (GTG) is formulated in the language of Geometric Calculus
(GC), initially discovered by Clifford and further developed by
Hestenes \cite{Hestenes}, Doran, Lasenby \cite{Doran}. GTG is
fully compliant with all classical tests of the standard
formulation of General Relativity (GR). However it is not based on
the principle of equivalence. Gauge symmetry plays a more
fundamental role in the theory than the spacetime metric. The
following two gauge principles for gravitation form the base of
the theory:

\begin{enumerate}
\item The Displacement Gauge Principle (DGP), which states that the
equations of physics must be invariant under arbitrary smooth
remappings of events onto spacetime.
\item The Rotation Gauge Principle (RGP), which formulates that the
equations of physics must be covariant under local Lorentz
rotations.
\end{enumerate}

DGP is a vast generalization of "\emph{translational invariance}"
in special relativity, so it has a comparable physical
interpretation. Accordingly, the DGP can be interpreted as
asserting that "\emph{spacetime is globally homogeneous}". In
other words, with respect to the equations of physics all
spacetime points are equivalent. DGP throws new light on
Einstein's Principle of General Covariance (PGC). The problem with
the PGC, as we saw above, is that it is not a true symmetry
principle \cite{Weinberg}. For a transformation group to be a
physical symmetry group, there must be a well defined "geometric
object" that the group leaves invariant. Thus the
"\emph{displacement group}" of the DGP is a symmetry group,
because it leaves the flat spacetime background invariant.
Following Noether's theorem, homogeneity of spacetime is
associated with the conservation of 4-linear momentum.

In special relativity, Lorentz transformations are passive
rotations expressing equivalence of physics with respect to
different inertial reference frames. In RGP, however, covariance
under active rotations expresses local physical equivalence of
different directions in spacetime. In other words, \emph{RGP
asserts that spacetime is locally isotropic}. Thus "passive
equivalence" is an equivalence of observers, while "active
equivalence" is an equivalence of states. Noether's theorem
establishes the conservation of the 4-angular momentum from the
local spacetime isotropy.

Therefore a violation of gauge symmetry in general relativity is
associated with a violation of energy-momentum conservation, which
naturally takes place in a non-homogeneous and anisotropic
universe! This is indeed what tends to be confirmed by the latest
cosmological observations \cite{carroll}, which would also
implicitly confirm the validity of Einstein-Maxwell-Proca
equations.

\end{document}